\documentclass[a4paper]{llncs}

\usepackage{graphicx}
\usepackage[dvipdfmx]{hyperref}

\begin{document}
\title{Documentation Generator Focusing on \\Symbols for the HTML-ized Mizar Library\thanks{The final publication is available at http://link.springer.com.}}
\author{Kazuhisa Nakasho\inst{1} \and Yasunari Shidama\inst{2}}
\institute{Shinshu University, Japan, \email{13st205f@shinshu-u.ac.jp}
\and Shinshu University, Japan, \email{shidama@cs.shinshu-u.ac.jp}}
\maketitle

\begin{abstract}
The purpose of this project is to collect symbol information in the Mizar Mathematical Library and manipulate it into practical and organized documentation. Inspired by the MathWiki project and API reference systems for computer programs, we developed a documentation generator focusing on symbols for the HTML-ized Mizar library. The system has several helpful features, including a symbol list, incremental search, and a referrer list. It targets those who use proof assistance systems, the volume of whose libraries has been rapidly increasing year by year.
\end{abstract}

\section{Motivation}
In mathematical knowledge management (MKM), expanding of the fields covered by formal methods has led to the rapid growth of formal mathematical libraries. For instance, the Mizar Mathematical Library (MML)\cite{Grabowski:2010,Matuszewski:2005,Naumowicz:2009} has grown to more than 2.7 million lines in 2015, and it has been increasing by approximately 0.1 million lines per year.

The development of formal mathematical libraries facilitates the reuse of mathematical symbols and theorems, thereby improving the efficiency of writing formal proofs. However, the increased volume of the libraries makes it difficult for users to grasp what and where symbols and theorems are defined. In recent years, developers of formal proofs have spent considerable time on search tasks in large-scale libraries, thereby decreasing the productivity of formal verification. Therefore, searching and browsing efficiency in large-scale libraries has been a crucial issue in MKM.

\section{Survey and Design Decision}
We analyze some existing tools for searching and browsing the Mizar library. The HTML-ized Mizar library\cite{Alama:2011a,Urban:2010} is one of the most successful documentation tools for formal mathematical libraries. The HTML-linked MML was first developed in the late 1990s for the former "Journal of Formalized Mathematics" \footnote{\url{http://mizar.org/JFM}} and then re-implemented by Dr. Josef Urban using the XML/XSLT technology. This new HTML-ization was also used in the MathWiki Project.\footnote{\url{http://www.ru.nl/foundations/research/projects/mathwiki/}} The system is capable of intuitive and rapid browsing as a result of hyperlinks being embedded into symbols, enabling users to jump from symbol occurrences to their definitions by clicking them. The system has been widely used by Mizar users because of its effectiveness and user-friendly design. However, because this system does not have retrieval functions, users are frequently obliged to grep symbols in the MML using text editors. Moreover, although the hyperlinks allow users to jump to their definitions, it is still difficult to, inversely, enumerate the symbols that include a particular symbol in their definitions.

MML Query\cite{Bancerek:2003,Bancerek:2004} is the most flexible and sophisticated search system for the MML. This system has its own query language, and users can input more detailed information regarding search objects than is possible using grep. However, users must learn and master the query language, thus this is a burden for beginners.

Conversely, in software development, most widely used programming languages have several types of API documentation generators, and almost all of the widely used libraries have their own online API documentation systems. Those API reference systems have common features, such as incremental search and a list of symbols that is automatically created by API documentation generators during library updates. Many documentation generators, such as Doxygen\footnote{\url{http://www.doxygen.org/}} and RDoc\footnote{\url{https://github.com/rdoc/rdoc}}, have contributed to the acceleration of software development.

We apply the software development approach to developing a documentation generator that works on the MML in order to overcome the drawback of existing search and browsing systems.

\section{Application}
Using the programming language Python,\footnote{\url{https://www.python.org/}} we developed a documentation generator\footnote{\url{https://github.com/aabaa/mmlfrontend}} that comprises the following three steps:
\begin{enumerate}
\item Parse the HTML-ized MML and collect symbols and their mutual relationships.
\item Clean and arrange those data.
\item Output reference documents in HTML format. Each file corresponds to one symbol.
\end{enumerate}
These steps take only a few minutes in total.\footnote{Windows 7, CPU: AMD A10-5800K 3.8 GHz (4-core), Memory: 16.0 GB}

\begin{figure}
\centering
\includegraphics[height=6.0cm]{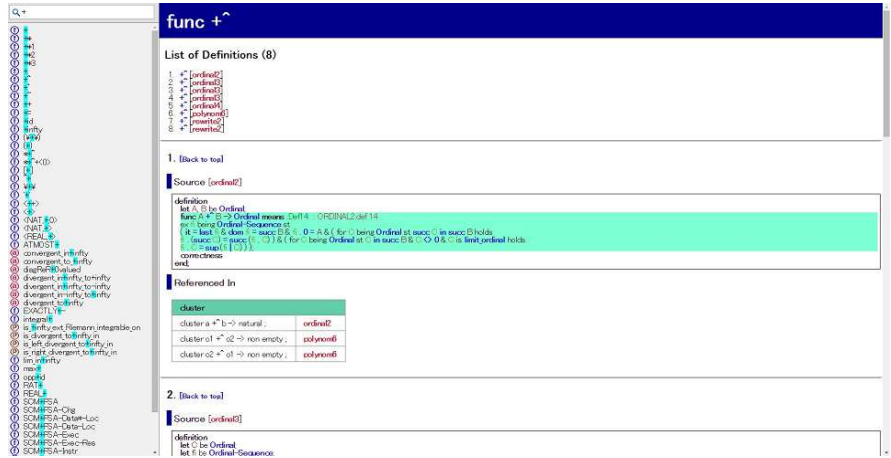}
\caption{Screenshot of the reference system.}
\label{fig:screenshot}
\end{figure}

The latest reference system produced by the generator is available at a website.\footnote{\url{http://webmizar.cs.shinshu-u.ac.jp/mmlfe/current/}} Fig.~\ref{fig:screenshot} shows a screenshot of the system.

The reference system offers the following helpful features:

{\bfseries Symbol List:}
There are nearly 9,000 symbols (predicate, mode, structure, functor, and attribute) in the MML, all of which are listed in the left pane of the system. The type of each symbol can be distinguished by the icon next to the symbol. Clicking a symbol in the list causes the corresponding page to be loaded into the main frame in left pane.

{\bfseries Incremental Search:}
An incremental search function is located at the top of the left pane. When several search words separated by blanks are input, the system combines the symbol list into symbols that contain all of the indicated words. As the system has an original search table, the function returns search results immediately. Users can quickly look up symbols defined in the MML, even without knowing the correct spelling.

{\bfseries Source Code:}
The symbol definition source code is imported from the HTML-ized MML. Symbols in bold font are hyperlinked to their definitions. Internal links pointing to their definitions in this reference system are in blue. External links pointing to their definitions in the original HTML-ized MML are in red.

{\bfseries Referrer List:}
Although the HTML-ized MML enables users to jump from symbol occurrences to their definitions by clicking them, it does not have a function to enumerate symbols that are used in the definitions of particular symbols. The new system organizes the list of referrers for each symbol, and users can check them easily.

\section{Conclusion and Future Work}
We utilized the API documentation technique from the field of software development to develop a new documentation generator that works on the MML. This system enables users to retrieve symbols quickly and intuitively using an incremental search function. Furthermore, users can easily check the types of symbols allowed to be used together by referrer lists. These functions have contributed considerably to improving the efficiency of formal proof development, and the system has gained a good reputation among the Mizar community. Additionally, the approach of the system is not specific to Mizar and the MML, thus all formal libraries would benefit from such a system. Therefore, the future versions of the system should support other formal languages and libraries.

We mention three remaining issues regarding the system:

{\bfseries Reimplementation with the XML-ized MML:}
The current documentation generator parses the HTML-ized MML instead of the XML-ized Mizar\cite{Urban:2005}. This is because the former represents relationships between symbols and their definitions as embedded hyperlinks, whereas it is difficult to collect these relationships from the latter. However, the extra process required to generate the HTML-ized MML takes considerable time. Therefore, we would like to change the system to work with the XML-ized MML in the future.

{\bfseries Theorem Search:}
A theorem search system requires semantic analysis, and machine learning would be a promising approach. Because this research is underway for automated reasoning\cite{Alama:2014,Urban:2006b}, we would like to apply the technique to an interactive search engine.

{\bfseries Tagged Comments:}
In software development, most documentation generators collect tagged comments, such as authors, purposes, and usages, and reflect them in API documents, whereas the current Mizar library does not have any tagged comments. Although Mizar is a comparatively readable formal language, it is sometimes difficult to discern a writer's intention from a source code. Consequently, such a function would work beneficially, if it were implemented. Furthermore, there is no standard for tagged comments in formal libraries, such a format should be developed in future work and then adopted by all formal libraries. 

We also suggest a possible application of the system:

{\bfseries Code Completion:}
Other major proof assistants have developed graphical interfaces, such as the jEdit plugins for Coq and Isabelle \cite{Tankink:2014,Wenzel:2012,Wenzel:2013}. Although the Mizar system provides an Emacs plugin\cite{Urban:2006a}, some users hope that a newer one will be offered on a modern integrated development environment (IDE). The incremental function of the system would assist in implementation of code completion for those IDE systems.

\section{Acknowledgment}
The authors wish to thank the members of the MathWiki Project for their preceding work.
Our research is deeply dependent on their product.
We especially express our gratitude to Dr. Josef Urban,
who is well known as a member of the MathWiki Project, for giving us some beneficial advice for the study.
We would also like to thank to Dr. Adam Naumowicz for helping us improve our paper.

\end{document}